\documentclass[sn-aps]{sn-jnl}

\jyear{2021}%

\theoremstyle{thmstyleone}%
\newtheorem{theorem}{Theorem}
\newtheorem{proposition}[theorem]{Proposition}%

\theoremstyle{thmstyletwo}%
\newtheorem{example}{Example}%
\newtheorem{remark}{Remark}%

\theoremstyle{thmstylethree}%
\newtheorem{definition}{Definition}%

\begin{document}

\title[Thermovoltaic Effects of van der Waals Heterojunctions based on Inert Conductor/Solution Interfaces]{Thermovoltaic Effects of van der Waals Heterojunctions based on Inert Conductor/Solution Interfaces}


\author*[1]{\fnm{Zhengliang} \sur{Wang}}\email{wzl111@zwu.edu.cn}

\author*[2]{\fnm{Gelin} \sur{Wang}}\email{wanggelin@nimte.ac.cn}


\affil[1]{\orgdiv{Zhejiang Wanli University},  \city{Ningbo}, \postcode{315100}, \state{Zhejiang}, \country{P.R.China}}

\affil[2]{\orgdiv{University of Nottingham Ningbo China (UNNC)},  \city{Ningbo}, \postcode{315100}, \state{Zhejiang}, \country{P.R.China}}



\abstract{It is found that if the inert conductor P has a larger electron work function $\phi$ and surface state function G than the inert conductor N, the inert conductor P and the inert conductor N are isolated by a separator and then immersed in the solution S (abbreviation: inert conductorP$\mid$solutionS$\mid$inert conductorN, or as P$\mid$S$\mid$N). Excluding the electrochemical reaction and thermoelectric effect of P$\mid$S$\mid$N, etc., it is measured that the voltage between the two conductors after the open circuit continues to increase to a certain stable maximum value. Then, the current after the closed-circuit continues to decrease to a certain stable minimum value. Analysis of the structure and properties of P$\mid$S$\mid$N shows that the inert conductor/solution interface relies on physical adsorption to construct van der Waals heterojunctions and that two van der Waals heterojunctions of different potentials form a P-N junction for P$\mid$S$\mid$N. The negative ions move to the inert conductor P under the action of the built-in electric field of the P-N junction to form a high-potential adsorption electric double layer. Under the action of the built-in electric field of the P-N junction, the positive ions move to the inert conductor N to form a low-potential adsorbed electric double layer, and the thermal energy is converted into the electric field energy of the two adsorbed electric double layers. The inert conductors P and N have different potentials.This electric field energy is expressed in the outer circuit when the circuit is open, due to the joint action of the electron work function $\phi$ and the surface state function G, P$\mid$S$\mid$N generates a larger built-in electric field and obtains a larger voltage. When the circuit is closed, P$\mid$S$\mid$N only has the effect of the surface state function G, which produces a smaller built-in electric field, results in a smaller voltage and current. This Thermoelectric conversion phenomenon is called the thermovoltaic effect. The heterojunction at the inert conductor/solution interface is a dissipative structure in a non-equilibrium state, which can self-organize the disordered thermal motion charges into an ordered adsorption electric double-layer structure. The theoretical analysis is consistent with the experimental phenomenon. The environment is a great clean heat source that is not limited by time and space. Thus, the thermovoltaic effect causes P$\mid$S$\mid$N to have inexhaustible electric energy. Several products can be developed such as active electric double-layer capacitors, permanent electrets, non-thermocouple thermocouples, thermovoltaic batteries, etc.}




\maketitle

\section{Introduction}\label{sec1}
In our experiments, when the high-purity inert electrodes Pt and Au were immersed in ultrapure H$_{2}$O (abbreviated as Pt$\mid$H$_{2}$O$\mid$Au), the maximum open-circuit voltage between the two electrodes was measured to be greater than 100mV, which was initially judged to be caused by the primary cells. However, we placed the Pt$\mid$H$_{2}$O$\mid$Au statically in a closed metal box to eliminate the interference such as the electrokinetic, thermoelectric, photoelectric, electromagnetic effects and concentration cell. At the same time, the Pt$\mid$H$_{2}$O$\mid$Au was short-circuited for one year to eliminate the electrochemical interferences of trace impurities in the electrode. Then conduct the test once a week during the year, each time the maximum open-circuit voltage between the two electrodes of Pt$\mid$H$_{2}$O$\mid$Au was measured to be greater than 100mV. This "never disappears" voltage is incredible, but the experiment objectively makes us suspect that Pt$\mid$H$_{2}$O$\mid$Au implies an unknown physical effect.

There is an internal potential difference at the contact interface of any two different objects, forming a built-in electric field, producing a contact electric double layer and constructing heterojunctions. The contact electric double layer has electric field energy, so the forming and changing process of the contact electric double layer is also a process of converting other energy into electric field energy. For example, a built-in electric field is formed at the contact interface between semiconductor P and semiconductor N, resulting in a contact electric double layer. This heterojunction is called a semiconductor P-N junction. When light shines on the semiconductor P-N junction, new hole-electron pairs are formed at the interface. Under the action of the electric field, the holes flow from the semiconductor N to the semiconductor P, and the electrons flow from the semiconductor P to the semiconductor N, which changes the size of the contact electric double layer. After the circuit is connected, a current is formed, which is the photovoltaic effect \cite{bi1}.

Comparing Pt$\mid$H$_{2}$O$\mid$Au with semiconductor P-N junction, water or solution is like liquid intrinsic semiconductor, while immersion of inert conductor P and N is like doping of intrinsic semiconducting. Positive and negative ions are like holes and electrons. The contact interface between the inert conductors P and N and the solution S forms a P-S junction and an S-N junction, which is synthesized as a P-N junction. The water molecules or electrolyte molecules in the solution spontaneously dissociate into positive ions and negative ions. Under the action of the built-in electric field of the P-N junction, the negative ions and positive ions move to the inert conductors P and N, forming two contact electric double layers with high and low potentials, and their potential difference is the voltage of Pt$\mid$H$_{2}$O$\mid$Au.

Therefore, this paper explores the structure and properties of P$\mid$S$\mid$N heterojunctions based on experiments to reveal P$\mid$S$\mid$N's voltage formation mechanism and expand the applications of P$\mid$S$\mid$N.

\section{Experiments}\label{sec2}
\subsection{Reagents and instruments}
\begin{enumerate}[1)]
	\item Inert conductors P and N are electronic conductors: typical inert conductors: Pt ($\geq$ 99.99\%), Au ($\geq$ 99.9\%), spectral graphite C ($\geq$ 99.99\%), graphite felt C ($\geq$ 98 \%), Surface oxidized metal electrode Ni ($\geq$ 99.99\%).
	
	\item Solution S is an ion conductor, H$_{2}$O ($\rho\textgreater$ 18.0 M$\Omega\cdot$cm), and the molar concentration of NaOH solution is c=0.1mol/dm$^{3}$.
	
	\item Containers: $\phi$40×150mm flat-bottomed quartz test tube, $\phi$60×150mm polypropylene bottle.
	
	\item Instruments: Electrochemical workstation (CHI660D), micro-current measuring instrument (measurement range: $19.99×10^{-12}$A-$19.99×10^{-6}$A), etc.
\end{enumerate}
\subsection{Experimental devices}
\begin{figure}[h]%
	\centering
	\includegraphics[width=0.7\textwidth]{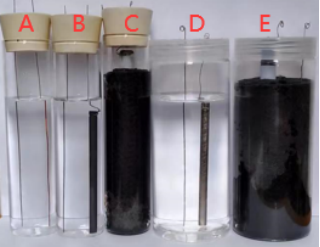}
	\caption{Experimental setups with five groups}\label{fig6}
\end{figure}

Inert conductors do not participate in chemical reactions and do not undergo any phase transitions in solution, but ions in the solution within the range of tunneling will exchange electrons with the energy band of the solid surface to transfer electrons \cite{bi2,bi3}. The two inert conductors need to be separated by an isolation film. The isolation film should meet the highest possible ionic conductance and the lowest possible electronic conductance, which can be omitted in water and dilute solutions. Fig \ref{fig6} of five groups shows representative P$\mid$S$\mid$N experimental setups: Group A: Pt$\mid$H$_{2}$O $\mid$ Au formed by $\phi$ 1 mm Pt wire and $\phi$ 1 mm Au wire. Group B: $\phi$ 1 mm Pt wire and $\phi$ 6 mm spectral graphite rod form Pt$\mid$H$_{2}$O$\mid$C, and the lead wires are all Pt wires. Group C: A layer of isolation film is wrapped around a $\phi$ 12mm MoSi$_{2}$ rod, and then graphite felt C is wrapped around it. The lead wires are all Ni wires to form C$\mid$H$_{2}$O$\mid$MoSi$_{2}$. Group D: $\phi$ 1 mm Pt wire and $\phi$ 6 mm Ni rod form Pt$\mid$NaOH(c)$\mid$Ni, and the lead wires are Pt wire and Ni wire respectively. Group E: a layer of isolation film is wrapped around a $\phi$ 12mm MoSi$_{2}$ rod, and then graphite felt C is wrapped around it, where the lead wires are all Ni wires to form C$\mid$NaOH(c)$\mid$MoSi$_{2}$.

\section{Experimental results and discussion}
\subsection{State quantity of P$\mid$S$\mid$N}
\begin{table}[h]
	\begin{center}
		\begin{minipage}{174pt}
			\caption{State quantity of P$\mid$S$\mid$N}\label{tab1}%
			\begin{tabular}{@{}llll@{}}
				\toprule
				No. & Device  & U(V) & I(A)\\
				\midrule
				A    & Pt$\mid$H$_{2}$O$\mid$Au   & 0.118  & 5.4x10$^{-10}$  \\
				B    & Pt$\mid$H$_{2}$O$\mid$C   & 0.129  & 5.1x10$^{-8}$   \\
				C    & C$\mid$H$_{2}$O$\mid$MoSi$_{2}$    & 0.350  & 6.7x10$^{-4}$  \\
				D    & Pt$\mid$NaOH(c)$\mid$Ni   & 0.105  & 3.68x10$^{-8}$  \\
				E    & C$\mid$NaOH(c)$\mid$MoSi$_{2}$   & 0.825  & 3.68x10$^{-3}$  \\
				
				\botrule
			\end{tabular}
			\footnotetext{maximum voltage of P$\mid$S$\mid$N after open circuit and minimum current after closed circuit (atmospheric pressure, room temperature 25 $^{\circ}$C)}
		
		\end{minipage}
	\end{center}
\end{table}
\begin{enumerate}[1)]
	\item Taking Group A and Group D as examples, the two electrodes of each group were short-circuited for 1 year, and the solution was detected by GFAAS, FAAS and ICP-AES. No Au ions and Ni ions were found, so the electrical energy of P$\mid$S$\mid$N does not come from electrochemical reactions.
	
	\item P$\mid$S$\mid$N is placed in a closed metal container and measured at a room temperature of 25°C without temperature changes. The maximum open-circuit voltage and minimum closed-circuit current between the two electrodes of P$\mid$S$\mid$N can still be measured. Therefore, the electrical energy excluding P$\mid$S$\mid$N comes from the electrokinetic effect, electromagnetic induction, photoelectric effect,  thermoelectric effects and concentration cell.
	
	\item Ambient heat energy is a huge heat source. P$\mid$S$\mid$N spontaneously absorbs ambient heat energy and causes a very small change in ambient temperature. P$\mid$S$\mid$N can only show continuous voltage when the ambient heat energy absorbed spontaneously is converted into electric energy. The thermovoltaic effect of P$\mid$S$\mid$N can reasonably explain this experimental phenomenon.
	
	\item The performance of P$\mid$S$\mid$N is related to the properties of inert conductor materials. For example, Pt has the largest positive potential, MoSi2 ceramics contain a large number of hydroxyl groups (-OH) and have the largest negative potential, porous inert conductors with a high specific surface area can obtain higher current (graphite felt C). The performance of P$\mid$S$\mid$N is related to the solution composition. For example, group E has a larger voltage and current than group C. There are many factors affecting the performance of P$\mid$S$\mid$N, so the experimental data in this paper applies only to qualitative analysis.
	
	\item P$\mid$S$\mid$N presents a continuous voltage and current. According to this experimental phenomenon, new types of permanent electrets with continuous voltage and active sensing devices with continuous self-powering can be developed.
\end{enumerate}

\subsection{Process quantity of P$\mid$S$\mid$N}

\begin{enumerate}[1)]
	\begin{figure}[h]%
		\centering
		\includegraphics[width=0.5\textwidth]{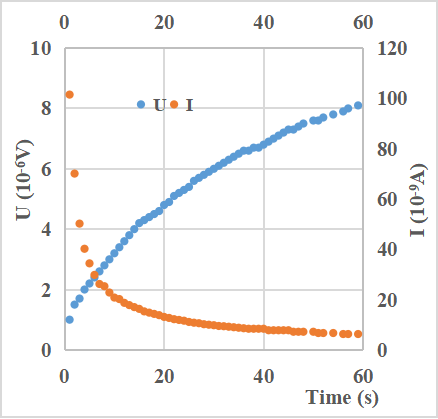}
		\caption{the relationship between the open-circuit voltage and closed-circuit current of group A and time.}\label{fig1}
	\end{figure}
	
	\item Fig.\ref{fig1} takes Group A as an example, after switching from a closed circuit to an open circuit the voltage continues to increase until it reaches a stable maximum value. The current from the open circuit to the closed-circuit continues to decrease until it reaches a stable minimum value, which remains the same for repeated cycle tests. The experimental phenomenon is similar to an RC circuit and is similar to the charging and discharging of a capacitor. Accordingly, an active electric double-layer capacitor integrating spontaneous conversion and storage of electrical energy can be developed. 
	
	\begin{figure}[h]%
		\centering
		\includegraphics[width=0.5\textwidth]{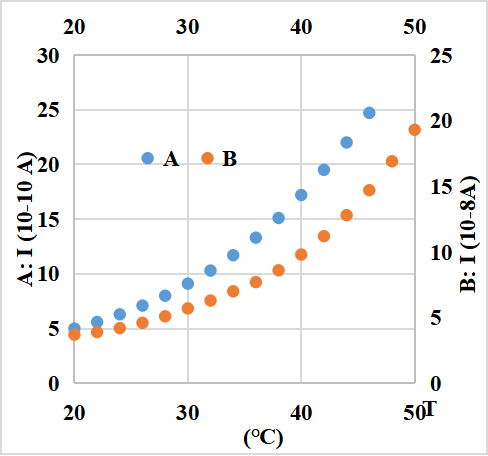}
		\caption{Relationship between closed-circuit current and temperature of groups A and B.}\label{fig2}
	\end{figure}
	\item Fig.\ref{fig2} takes groups A and B as examples. At a certain temperature, there is a stable continuous minimum current after P$\mid$S$\mid$N is closed. This current increases with the increase of temperature while the current is a single-valued function of temperature. In this way, the active temperature sensors such as non-temperature differential thermocouples can be developed.
	\begin{figure}[h]%
		\centering
		\includegraphics[width=0.5\textwidth]{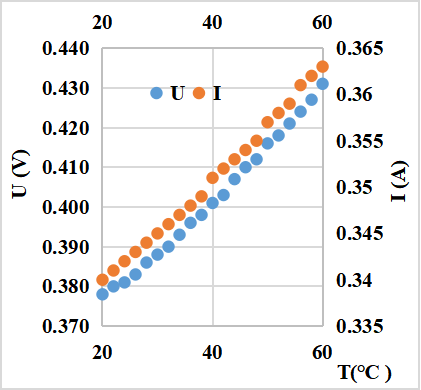}
		\caption{ Relationship between voltage, current, and temperature after group E is externally connected with a 990 $\Omega$ resistor.}\label{fig3}
	\end{figure}
	\item Fig.\ref{fig3} takes the external 990$\Omega$ resistor of group E as an example. The continuous output power at room temperature is small, and the output power is related to the temperature in quadratic linear equation. The energy conversion efficiency can be improved by increasing the temperature. In this way a sustainable self-sufficient power source can be developed.
\end{enumerate}

\section{Theoretical analysis}
\subsection{Applying the principle of physical adsorption to reveal the P$\mid$S$\mid$N energy conversion mechanism}
Physical adsorption, also known as van der Waals adsorption, is caused by intermolecular forces between the adsorbent and the adsorbent, which are also known as van der Waals forces. Water is a strongly polar molecule, and inert conductors will adsorb water molecules in water in a characteristic orientation, forming a water dipole electric double layer at the inert conductor/water interface. The adsorption phenomenon in which the same dipole molecules or ions of the same sign are gathered to the interface region is called characteristic adsorption. Any substance that can adsorb at the electrode/solution interface and reduce the interfacial tension is called a surface-active substance. Most inorganic anions are surface-active species and have typical ion adsorption laws. The anion squeezes out the water molecules on the surface of the inert conductor, and the anion characteristic adsorption occurs. If the characteristic adsorption of anions occurs in the inert conductor/solution, an electric double layer is formed in the solution due to the electrostatic interaction between the characteristically adsorbed anions and the cations in the solution, which is called the adsorption electric double-layer \cite{bi4}. The formation of the adsorbed electric double layer is a spontaneous process, in which thermal energy is converted into the electric field energy of the adsorbed electric double layer. This inert conductor/solution interface is therefore referred to in this paper as a van der Waals heterojunction. Characteristic adsorption in physical adsorption plays one of the key roles in the P$\mid$S$\mid$N thermoelectric conversion mechanism.
\begin{figure}[h]%
	\centering
	\includegraphics[width=0.9\textwidth]{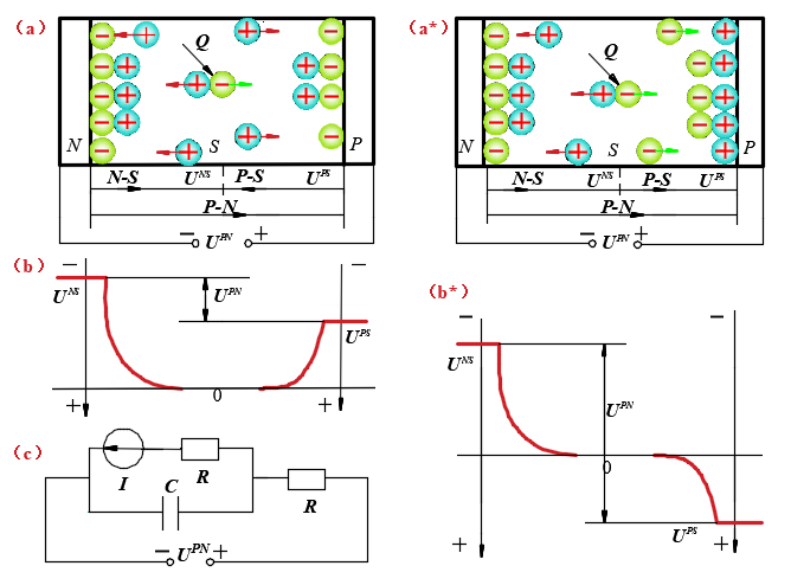}
	\caption{P$\mid$S$\mid$N working principle diagram: (a)the adsorption of similar charge P$\mid$S$\mid$N diagram; (b)corresponding potential diagram; (a$^{*}$)adsorption of heterogeneous charge P$\mid$S$\mid$N diagram;(b$^{*}$)corresponding potential diagram. (c)P$\mid$S$\mid$N equivalent circuit diagram.} \label{fig4}
\end{figure} 

Compared with the inert conductor N, the inert conductor P is prone to adsorption of anion characteristics. In other words, the inert conductor N is prone to adsorption of cation characteristics compared to the inert conductor P. Fig.\ref{fig4}(a), (b) shows the P$\mid$S$\mid$N absorbing the same type of charge inert conductor combination (such as Pt$\mid$H2O$\mid$Au). Fig.\ref{fig4}(a*) and (b*) show the P$\mid$S$\mid$N absorbing heterogeneous charge inert conductor combination (such as C$\mid$H$_{2}$O$\mid$MoSi$_{2}$).

The contact interface of the two materials mostly exchanges charges, and the contact electric double layer exaggerates the two phases. However, since there is no charge exchange at the inert conductor/solution interface, the polar molecules and ions in the solution are aligned on the solution side of the interface, so the contact electric double layer only appears in the solution phase \cite{bi5,bi6,bi7}. With the contacting electric double layer to be the adsorption of the electric double layer, the inert conductor is like the electrode of the electric double layer capacitor, and the potential of the adsorbed electric double layer becomes the surface potential of the inert conductor. The inert conductors P and N have different potentials, so the potential difference between the two adsorbed electric double layers is the P$\mid$S$\mid$N voltage. As shown in Fig.\ref{fig4}(b) and (b*),

\begin{equation}
	U^{P N}=U^{P S}-U^{N S}
\end{equation}
Obviously, (b*) in Fig.\ref{fig4} can obtain a larger voltage $U^{PN}$ than (b), such as C$\mid$H$_{2}$O$\mid$MoSi$_{2}$ is the preferred combination. The P$\mid$S$\mid$N equivalent circuit is shown in Fig.\ref{fig4}(c), which can be regarded as a self-powered electric double-layer capacitor. The theoretical analysis is consistent with the experiment.

The potential of the van der Waals heterojunction at the inert conductor directly presenting the inert conductor/solution interface also plays an essential role in the P$\mid$S$\mid$N thermoelectric conversion mechanism.

\subsection{Invoking the theory of heterojunctions to reveal the P$\mid$S$\mid$N energy conversion mechanism}
As shown in Fig. \ref{fig4}: Since any two different objects have an internal potential difference (also known as Galvani potential difference) while contacting, the potential differences between the P-S junction and N-S are -$U^{PS}$ and $U^{NS}$. P-N junction has an internal potential difference of -$U^{PN}$. The internal potential difference and the external potential difference (also called voltaic potential difference) are equal in magnitude and opposite in direction. The negative ions move to the inert conductor P under the action of -$U^{PS}$ to form a contact electric double layer, which has a higher external potential difference $U^{PS}$. The positive ions move to the inert conductor N under the action of -$U^{PN}$ to form a contact electric double layer, which is composed of lower external potential UNS. Thus the external potential difference $U^{PN}$ of the P-N junction is also the voltage $U^{PN}$ of P$\mid$S$\mid$N.

Electronic work function $\phi$ and surface state function G are two important physical quantities to measure the surface properties of materials. The size of the electron work function $\phi$ indicates the strength of the electron binding in the material. The surface state function G is the ability of the system to do non-volume work under isothermal and isobaric conditions. The free energy of the inert conductor/solution interface is reduced by $\Delta$G, so the adsorption process can be realized \cite{bi8}. If the amount of charge in contact with the electric double layer is q:

\begin{equation}
\begin{gathered}
	U^{p s}=\left(\Delta \phi^{p s}+\Delta G^{p s}\right) / q \\
	U^{S N}=\left(\Delta \phi^{S N}+\Delta G^{S N}\right) / q
\end{gathered}
\end{equation}

Then the voltage $U_{o}^{PN}$ of P$\mid$S$\mid$N in open circuit:\\
\begin{equation}
	U_{o}^{P N}=\left(\Delta \phi^{P N}+\Delta G^{P N}\right) / q
\end{equation}
When P$\mid$S$\mid$N is closed:\\
\begin{equation}
	\Delta \phi=\Delta \phi^{P S}-\Delta \phi^{S N}-\Delta \phi^{N P}=0
\end{equation}
Then the voltage $U_{c}^{PN}$ of P$\mid$S$\mid$N in the closed-circuit:\\
\begin{equation}
	U_{c}^{P N}=\Delta G^{P N} / q
\end{equation}
So:\\
\begin{equation}
	U_{c}^{P N} \leq U_{o}^{P N}
\end{equation}
The voltage of P$\mid$S$\mid$N in the open circuit comes from the combined action of the electron work function $\phi$ and the surface state function G, and the voltage in the closed circuit only comes from the action of the surface state function G. Therefore, when P$\mid$S$\mid$N switches from closed circuit to open circuit, the voltage continuously increases from $U_{c}^{PN}$ to a stable maximum value $U_{o}^{PN}$, and the open-circuit state has a continuous voltage $U_{o}^{PN}$. From open circuit to closed circuit, the voltage continuously decreases from $U_{o}^{PN}$ to a stable minimum value of $U_{c}^{PN}$, and the closed-circuit state has a continuous electromotive force of $U_{c}^{PN}$.  The theoretical analysis is consistent with the experimental phenomenon.

Therefore, van der Waals heterojunctions of electron/ion conductors such as P$\mid$S$\mid$N have a unique mechanism for thermoelectric conversion.

\subsection{Invoking non-equilibrium thermodynamics to argue for a P$\mid$S$\mid$N energy conversion mechanism}
The inert conductor/solution interface is an interphase transition zone with different properties from the two-phase matrix. The P$\mid$S$\mid$N closed loop has a built-in electric field generated by ($\Delta G^{PS}-\Delta G^{NS}$), so P$\mid$S$\mid$N is a non-equilibrium state system which does not belong to the category of equilibrium thermodynamics. Therefore, the thermovoltaic effect of P$\mid$S$\mid$N does not violate the second law of thermodynamics, and P$\mid$S$\mid$N is not a perpetual motion machine of equilibrium thermodynamics.

The actual process of P$\mid$S$\mid$N is carried out in the atmospheric environment. The atmospheric environment is an enormous heat source. When it exchanges limited heat with the solution, its temperature and pressure changes are infinitely small, so the temperature T of P$\mid$S$\mid$N after endothermic or exothermic heat is treated as an invariant. According to the first law of thermodynamics, there are two sources of changing the internal energy dE of the solution, named the external work dW and the heat transfer dQ. In non-equilibrium thermodynamics, Prigogine divides the entropy of the solution into two parts. One part is generated by the irreversible process of thermal motion inside the solution, and this part of the entropy is called the entropy generation term, which is represented by S$_{i}$. The other part is caused by the interaction between the solution and the outside world. This part of the entropy is called the entropy flow term and is represented by S$_{e}$. The internal energy dE for the solution absorption dQ at equilibrium state are:

\begin{equation}
	\begin{aligned}
		&d E=d Q=T d S_{i} \\
		&d S_{i}=\frac{d Q}{T}=\frac{d E}{T} \geq 0
	\end{aligned}
\end{equation}
This process is in line with the principle of increasing entropy. The process of immersing the inert conductor P and the inert conductor N in the solution is the process of introducing the built-in electric field energy of $d(G^{PS}-G^{NS})$, and is also the process of doing non-volume work dW to the solution.

\begin{equation}
	d W=d\left(G^{P S}-G^{N S}\right)
\end{equation}

If the solution of P$\mid$S$\mid$N absorbs heat dQ, the internal energy dE is:

\begin{equation}
	d E=d Q+d\left(G^{P S}-G^{N S}\right)
\end{equation}

Then:

\begin{equation}
	\begin{aligned}
		&d Q=d E-d\left(G^{P S}-G^{N S}\right) \\
		&d S=\frac{d Q}{T}=\frac{d E}{T}-\frac{d\left(G^{P S}-G^{N S}\right)}{T}=d S_{i}-d S_{e}
	\end{aligned}
\end{equation}

If:
\begin{equation}
	\lvert d S_{i}\rvert \leq\lvert d S_{e}\rvert
\end{equation}
Then:
\begin{equation}
	d S \leq 0
\end{equation}

At the same time, the system can maintain some sort of orderly state as long as a sufficiently large negative entropy flow $dS_{e}$ is maintained. The built-in electric field energy $d(G^{PS}-G^{NS})$ plays a key role in the disorder-to-order organization process. Physical adsorption is a spontaneous and reversible process, and the van der Waals heterojunction produced by physical adsorption is a conversion of thermal energy into electric field energy in the dual electric layer. Therefore, P$\mid$S$\mid$N is a non-equilibrium dissipative structure. The self-organization of low-grade thermal energy into high-grade electrical energy follows non-equilibrium thermodynamics. The establishment of this ordered structure is developed by this open system itself, which is called the self-organization phenomenon \cite{bi9}. This physical phenomenon is consistent with the characteristics of the thermovoltaic effect.

\section{Conclusion}
\begin{enumerate}
	\item Through the long-term measurement and analysis of multiple groups of P$\mid$S$\mid$N experimental devices, it is confirmed by the exclusion method that the inexhaustible electrical energy of P$\mid$S$\mid$N comes from the ambient thermal energy absorbed spontaneously.
	
	\item P$\mid$S$\mid$N structure and properties are a): The inert conductor/solution interface relies on physical adsorption to construct van der Waals heterojunctions. b): The contact doublet at the inert conductor/solution interface occurs only in the solution phase. Thus the inert conductor directly shows the potential of the heterojunction. c): In an open circuit, P$\mid$S$\mid$N obtains a large voltage due to the combined action of the electron work function and the surface state function; in a closed circuit, P$\mid$S$\mid$N obtains a smaller voltage due to the action of the surface state function only. d): P$\mid$S$\mid$N is a non-equilibrium dissipative structure that can self-organize low-grade thermal energy into high-grade electrical energy, manifesting as the thermo-voltaic effect. 
	
	\item We invented the inert conductor P$\mid$, solution S$\mid$ and inert conductor N device with thermovolt effect, and discovered the physical phenomenon that the van der Waals heterojunction of the electronic conductor/ionic conductor can spontaneously absorb the environmental heat energy and convert it into the electric double layer electric field energy. It opens up a new field of non-equilibrium thermodynamics, provides a new way of thermoelectric conversion of self-organized thermal energy into electrical energy, obtains a sustainable, self-sufficient power source for micro/nano-systems, and can develop many components with special properties.
	
\end{enumerate}

\bibliography{sn-bibliography}

\begin{thebibliography}{1}
\providecommand{\url}[1]{{#1}}
\providecommand{\urlprefix}{URL }
\providecommand{\doi}[1]{\url{https://doi.org/#1}}
\bibcommenthead

\bibitem{bi1}
D.M. Chapin, C.S. Fuller, G.L. Pearson, A new silicon p-n junction photocell
  for converting solar radiation into electrical power.
\newblock Journal of applied physics \textbf{25}(5), 676--677 (1954)

\bibitem{bi2}
W.~Xu, H.~Zheng, Y.~Liu, X.~Zhou, C.~Zhang, Y.~Song, X.~Deng, M.~Leung,
  Z.~Yang, R.X. Xu, et~al., A droplet-based electricity generator with high
  instantaneous power density.
\newblock Nature \textbf{578}(7795), 392--396 (2020)

\bibitem{bi3}
B.H. Jin~Tan, Qunwei~Tang, Watervoltaic materials and energy conversion device
  based on carbon nanomaterials.
\newblock Chinese Science Bulletin \textbf{63}(27), 2818--2832 (2018)

\bibitem{bi4}
X.~Fufan, L.~Wenbin, \emph{Physical Chemistry} (Tianjin University Press,
  Tianjin, 2007)

\bibitem{bi5}
J.B.X. et~al., Power from water and graphene.
\newblock Chinese Science Bulletin \textbf{63}, 2806 – 2817 (2018)

\bibitem{bi6}
Z.~Yuan, C.~Pan, Quantifying electron-transfer in liquid-solid contact
  electrification.
\newblock Science Bulletin \textbf{65}(11), 868--869 (2020)

\bibitem{bi7}
J.~Nie, Z.~Ren, L.~Xu, S.~Lin, F.~Zhan, X.~Chen, Z.L. Wang, Probing
  contact-electrification-induced electron and ion transfers at a liquid--solid
  interface.
\newblock Advanced Materials \textbf{32}(2), 1905,696 (2020)

\bibitem{bi8}
L.~Di, \emph{Principles of Electrochemistry} (Beijing University of Aeronautics
  and Astronautics Press, Beijing, 2019)

\bibitem{bi9}
R.~Li, Non-equilibrium thermodynamics and dissipative structure.
\newblock Press of Tsinghua University, Beijing  (1986)

\end{thebibliography}


\end{document}